# High-throughput discovery of robust room-temperature superconductors among complex ternary clathrate hydrides

Tiancheng Ma[1], Decheng An[1], Zihan Zhang[1], Shuting Wu[1], Tian Cui[2,1,*], Defang Duan[1,**],

[1] *Key Laboratory of Material Simulation Methods & Software of Ministry of Education and State Key Laboratory of Superhard Materials, College of Physics, Jilin University, Changchun 130012, China*
[2] *Institute of High Pressure Physics, School of Physical Science and Technology, Ningbo University, Ningbo 315211, China*

*Corresponding author: cuitian@nbu.edu.cn
**Corresponding author: duandf@jlu.edu.cn

**Abstract**

After the decade-long exhaustive study of binary high-$T_c$ superconducting hydrides, the frontier of this stimulating research field has recently shifted to ternary hydrides with much expanded conformational space in search of coveted room-temperature superconductors. This task, however, presents a formidable challenge due to enormous demands on computational resources. Here, we devise an efficient high-throughput approach using keen material insights and a self-built database to screen for robust ternary hydrides in clathrate structures, which were proven to host highest $T_c$ in binary hydrides, and to estimate $T_c$ by a reliable empirical formula. This approach has made it possible to uncover a diverse set of complex multiple-hydrogen-cage ternary hydrides hosting near or above room-temperature $T_c$, which are beyond the reach of prevailing structure search methods. This study establishes a distinct paradigm that opens a fresh avenue to enable and accelerate the discovery of promising room-temperature superconductors among unprecedented complex clathrate hydrides.

Key words: room-temperature superconductor, high pressure, clathrate hydrides, high-throughput calculation.

Ashcroft predicted in 1968 that metallic hydrogen may be a room-temperature superconductor[1], but the pressure to drive the phase transition from insulating molecular hydrogen to metallic hydrogen turns out to exceed 500 GPa[2-4], which is prohibitively challenging to current experimental synthesis and characterization. To circumvent this difficulty, Ashcroft in 2004 proposed[5] to use chemical precompression in hydrogen-rich compounds to realize metallic hydrogen states. Following this idea, $H_3S$ with high superconducting transition temperature ($T_c$) above 200 K at 200 GPa was theoretical predicted and then experimental confirmed[6-8]; the search for room-temperature superconductivity in compressed hydrides ensued, leading to the discovery of a wide variety of superconducting binary hydrides. Currently known high-$T_c$ hydrides can be grouped into three categories: covalent (e.g., $H_3S$), clathrate (e.g., $LaH_{10}$[9,10]) and layered (e.g., $HfH_{10}$[11]). The clathrate hydrides are the most prevalent with the highest $T_c$ and become the focus of theoretical and experimental effort. The first theoretically predicted clathrate hydride was $CaH_6$[12], with the H atoms forming sodalite-like $H_{24}$ cages,[9] which was recently confirmed by experiments with measured $T_c$ of 210-215 K at 172-185 GPa[13,14]. Clathrate hydrides with higher H constant have been synthesized, e.g. $MH_{10}$ (M = La, Ce, Th)[15-18] containing $H_{32}$ cages and $MH_9$ (M = Y, Ce, Th)[17-21] stacked by $H_{29}$ cages, among which $LaH_{10}$ and $YH_9$ host $T_c$ of 250-260 K (190 GPa) and 243-262 K (182-201 GPa), respectively. The metal atoms in these binary clathrate hydrides have the same Wyckoff symbol and chemical environment so that they usually adopt single-cage clathrate structures containing only one type of H cage. The limited structural space has exhausted the binary clathrate hydrides with simple metal arrangements such as bcc ($CaH_6$), fcc ($LaH_{10}$) and dhcp ($YH_9$).

Recent studies of clathrate hydrides have turned to ternary hydrides for their richer stoichiometry and larger conformational space. Elemental substitution based on binary clathrate hydrides has thus far been the main route for making ternary clathrate hydrides such as $(La, Y)H_{10}$[22] and $(La, Ce)H_9$[23,24], with $T_c$ of 253 K (183 GPa) and 176 K (100 GPa), respectively; but their crystal structures takes single-cage forms isostructural to binary hydrides and, consequently, the phase space of ternary clathrate hydrides is still not fully explored. Recent synthesis of clathrate hydrides with stoichiometry $M_4H_{23}$ (M = Eu[25], Ba[26], La[27,28], Lu[29], Y[30]) offers a useful hint. Here, the metal atoms are arranged as an A15 structure with two unequal sites, and the H framework comprises distinct $H_{20}$ and $H_{24}$ cages, suggesting a new way to design ternary clathrate hydrides by stacking different types of H cages according to the metal template, forming multi-cage clathrate structures. Another clathrate compound $Li_2LaH_{17}$ ($Fd$-$3m$) follows this design with a structure comprising $H_{20}$ and $H_{28}$ cages[31]. The structure of hot-superconductor $Li_2MgH_{16}$ is analogous to that of $Li_2LaH_{17}$, although some H defects are present[32]. Additional ternary clathrate hydrides with multi-cage structures, such as $NaLi_3H_{23}$[33], $Li_2NaH_{17}$[33] and $LaSc_2H_{24}$[34], were recently predicted to be room-temperature superconductors. In contrast to binary clathrate hydrides, the non-integer H/metal ratios are common in ternary clathrate hydrides due to different types of H cages[35]. The complex structures of these ternary compounds make it hard for crystal structure search methods that were widely used in the study of binary hydrides. Furthermore, another difficulty comes from the need to find an effective yet still accurate way to determine $T_c$ of candidate compounds to evaluate their potential to be room-temperature superconductors. Methods need to be devised to overcome this dual obstacle.

In this work, we tackle the seemingly insurmountable challenge of searching for viable ternary clathrate hydrides with distinct H cages and also assessing their $T_c$. We proceed by noting two important clues. First, clathrate structures are common among hydrates, and water molecules ($H_2O$) adopt cage structures, each of which contains 20 to 36 $H_2O$, forming a rich variety of clathrate hydrates with multi-cage structures[36]. Second, group 14 elements tend to $sp^3$ hybridize in tetrahedral bonding pattern and form clathrate frameworks[37]. Based on these insights, we designed a series of geometrically rational clathrate hydrides with multi-cage structures in hydrates and frameworks of group 14 elements. For effective screening of superconductivity in identified clathrate hydrides, we use an empirical linear formula, known as the AE model, for estimating $T_c$. Our approach successfully found multiple viable ternary clathrate hydrides in Li-Na-H, Li-Ca-H, Li-Sr-H and Na-Sc-H systems with calculated $T_c$ near or above room temperature. These findings demonstrate the efficacy of our strategy for rational structure design guided by known structural arrangements as key building blocks combined with rapid $T_c$ evaluation via reliable empirical estimate, which produced promising results to guide experimental synthesis and characterization and also set stage for further exploration of complex hydrides.

## RESULTS

**The high-throughput workflow and the modified AE model**

For our designed ternary clathrate hydrides with multi-cage structures, we followed the naming rules of Roman numerals used in a recent study[33]. The high throughput (HTP) screening for their thermodynamic stability and superconductivity were performed using our self-developed FF7 (Fast Funnels with 7 functions) codes package[38]. The workflow is shown in Fig. 1a. We first perform HTP calculations for the thermodynamic stability of the ternary clathrate hydrides using a self-built hydrides database containing high-pressure phases of related simple substances and binary hydrides, which are main precursors for synthesizing ternary hydrides, and the formation enthalpy against them ($\Delta H$) can generally reflect the thermodynamic stability and provide useful guidance for experimental synthesis. We applied this method in the system of $XYH_8$, the structure of which is isostructural with $LaBeH_8$[39], to check its validity. It successfully predicted the thermodynamic stability of $LaBeH_8$, $LaBH_8$, $YBeH_8$, $CeBeH_8$, $ThBeH_8$ and $BaSiH_8$, which are in good agreement with previous results[39-43] (see Supplementary Fig. 1). This result suggests that the HTP method for thermodynamic stability based on our database and codes is effective and reliable. Moreover, for the thermodynamically stable and metastable hydrides with $\Delta H$ less than 35 meV/atom, we employed the VASP code to calculate their electronic band structures, density of states (DOS) and electron localization function (ELF). We then assessed $T_c$ of the thermodynamic stable and metastable hydrides in a HTP manner using an empirical linear formula AE model[44], which uses descriptors of the average value of ELF ($AE$) at the middle of H-H bonds in the cages and the contribution of H atoms to the total DOS at the Fermi level ($RH$). However, there are outliers in the prediction of $T_c$ of clathrate hydrides $Ca_3LiH_{23}$ and $CeLi_2H_{17}$ using the AE model. The calculated $T_c$ of $Ca_3LiH_{23}$[33] and $CeLi_2H_{17}$ are 51 K and 14 K, whereas the $T_c$ estimated by the AE model are 259 K and 151 K, respectively. These two hydrides share the same properties of high $RH$, low absolute values of H-DOS at Fermi level per H atom (*AH*) and low $T_c$. The AE model only considers the relative contribution of H-DOS to $N(E_f)$ and not its absolute value, resulting in less predictive capability for these hydrides. We define a penalty function

$$f(x) = \begin{cases} 40x & 0 < x < 0.025 \\ 1 & x \geq 0.025 \end{cases} \quad (1)$$

to integrate the variant *AH* into the AE model to modify it. The formula for estimated $T_c$ by the modified AE model ($T_c^{AE}$) is

$$T_c^{AE} = 896.7 \times f(AH) \times AE \times RH^{\frac{1}{3}} - 163.9 \quad (2)$$

and the predicted results are shown in Fig. 1b. Overall, most of the prediction errors of $T_c$ are within 40 K and the predicted $T_c$ values conform well to experimentally measured values. The modified AE model allows us to screen hydrides with high $T_c$ at a very small computational cost. Finally, the Quantum Espresso (QE) package[45] was employed to calculate phonon spectra and electron-phonon coupling (EPC) for the thermodynamically stable and metastable hydrides to screen for promising candidates with high $T_c$ values.

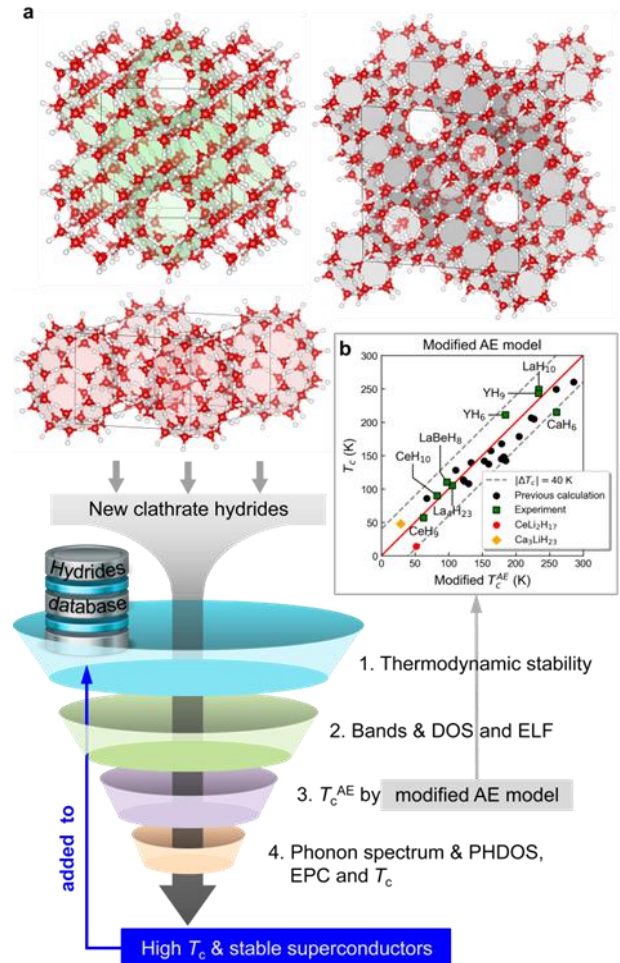

**Fig. 1 | High-throughput Workflow. a.** The high-throughput calculation flowchart. **b.** The scatter plots of directly calculated superconducting critical temperatures against the results by the modified AE model.

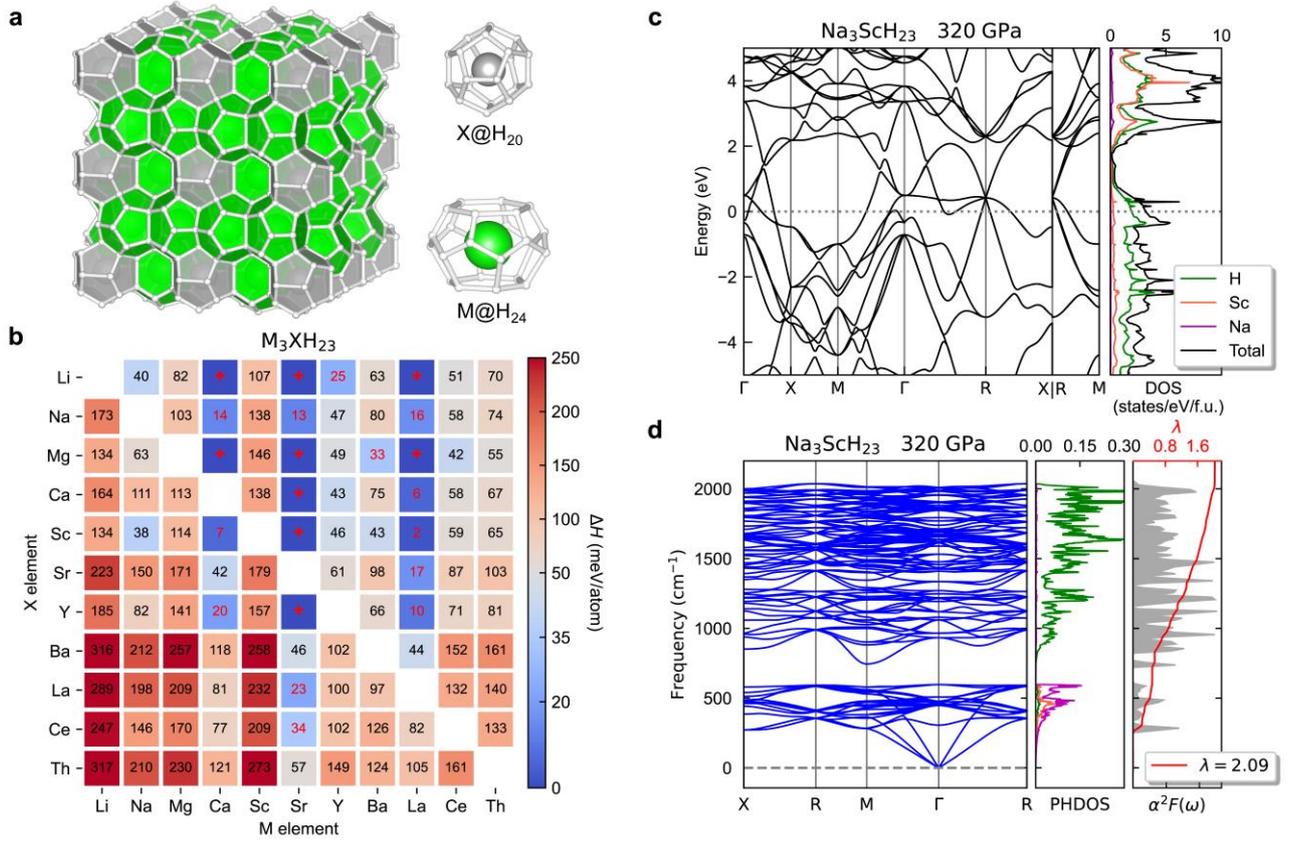

**Fig. 2 | Crystal structure, thermodynamic stability and electronic structures of type I clathrate hydrides.**
**a.** Structure and cage units of type I clathrates hydride $M_3XH_{23}$. **b.** Heatmap for thermodynamic stability of hydrides $M_3XH_{23}$ at 200 GPa. Colors/numbers in squares indicate $\Delta H$. Metastable hydrides ($\Delta H$<35 meV/atom) are marked by red numbers, while red crosses indicate thermodynamically stable hydrides. **c.** Electronic band structure (left) and density of states (right) of $Na_3ScH_{23}$ at 320 GPa. **d.** Phonon dispersion curves (left), phonon density of states (middle) and Eliashberg spectral function (right) of $Na_3ScH_{23}$ at 320 GPa.

**Type I clathrate hydrides**

Type I clathrate hydrides $M_3XH_{23}$ in $Pm\bar{3}n$ space group symmetry comprises two $H_{20}$ $[4^05^{12}6^0]$ cages and six $H_{24}$ $[4^05^{12}6^2]$ cages. The three superscripts in the brackets indicate the number of quadrilaterals, pentagons and hexagons that make up the cage unit. The number ratio of H to metal atoms is 5.75, slightly less than that of $CaH_6$. The metal atoms are arranged as in the A15 structure, as shown in Fig. 2a. In a recent theoretical study, $Na_3LiH_{23}$[33] was predicted to be a high-temperature superconductor with $T_c$ of 302-323 K at 320 GPa. We performed HTP calculations for thermodynamic stability of type I clathrate hydrides at 200 GPa and the results are summarized in the heatmap diagram given in Fig. 2b. We find eight hydrides, $Ca_3LiH_{23}$, $Ca_3MgH_{23}$, $Sr_3LiH_{23}$, $Sr_3MgH_{23}$, $Sr_3CaH_{23}$, $Sr_3YH_{23}$, $La_3LiH_{23}$ and $La_3MgH_{23}$, that are thermodynamically favorable at 200 GPa, and $T_c$ calculated by the self-consist solution of isotropic Eliashberg function ($T_c^E$) are 97 K (200 GPa), 223 K (200 GPa), 79 K (200 GPa), 166 K (200 GPa), 197 K (300 GPa), 184 K (300 GPa), 95 K (200 GPa) and 135 K (200 GPa), respectively. Remarkably, $T_c^{AE}$ of metastable hydride $Na_3ScH_{23}$ is 324 K, which inspired further research. We constructed the convex hull of the Na-Sc-H system (see Supplementary Fig. 2) by an extensive structure search of more than 40,000 structures. $Na_3ScH_{23}$ falls on the convex hull at 350 GPa, indicating its thermodynamic stability. Its $N(E_f)$ is dominated by electrons from H atoms as shown in Fig. 2c. We then calculated the phonon dispersion curves and Eliashberg spectral function to examine dynamic stability and superconductivity (Fig. 2d). The structure is dynamically stable at pressures above 320 GPa, and calculated EPC constant ($\lambda$) and $\omega_{log}$ are 2.09 and 1,390 K, respectively, and more than 74% of $\lambda$ is contributed by the optical phonon modes that are dominated by the vibrations of H atoms with frequency above 750 cm$^{-1}$. The $T_c$ calculated by Allen–Dynes modified McMillan ($T_c^{AD}$) is 250 K with $\mu^*$=1.0. $T_c^E$ reaches 289 K, which is in good accordance with the value by the modified AE model.

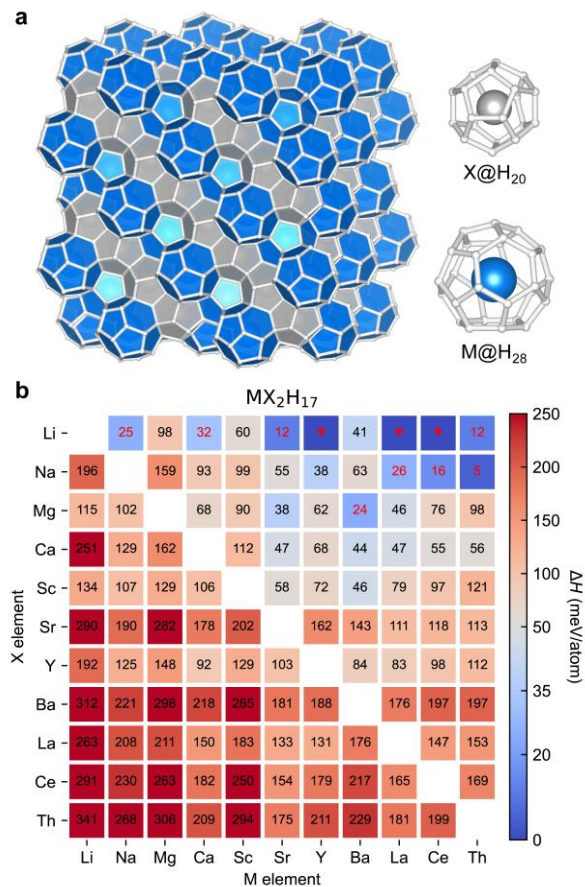

**Fig. 3 | Crystal structure and thermodynamic stability of type II clathrate hydrides. a.** Structure and cage units of type II clathrate hydrides $MX_2H_{17}$. **b.** Heatmap for thermodynamic stability of hydrides $MX_2H_{17}$ at 200 GPa. Colors/numbers in squares indicate $\Delta H$. Metastable hydrides ($\Delta H$<35 meV/atom) are marked by red numbers, while red crosses indicate thermodynamically stable hydrides.

**Type II clathrate hydrides**

Type II clathrate hydrides $MX_2H_{17}$ in $Fd\bar{3}m$ space group symmetry contains four $H_{20}$ [$4^05^{12}6^0$] cages and two $H_{28}$ [$4^05^{12}6^4$] cages (Fig. 3a). The ratio of number of H atoms to metal atoms is 5.67. This type of clathrate hydrides has been studied in the Li-R-H (R = Sc, Y, La) system[31], and $LaLi_2H_{17}$ and $YLi_2H_{17}$ were predicted to be thermodynamically stable at 300 GPa with $T_c$ of 108 K (200 GPa) and 156 K (160 GPa), respectively. Also, $NaLi_2H_{17}$[33] in type II clathrate structure was recently reported to be a possible thermodynamically stable room-temperature superconductor at 350 GPa with $T_c$ of 309-330 K. Thermodynamical stability of type II clathrate hydrides is summarized in Fig. 3b. Hydrides $LaLi_2H_{17}$ and $YLi_2H_{17}$ are thermodynamically stable at 200 GPa, in accordance with the previous results[31] and showing reliability of our database and HTP workflow. We also found a new hydride $CeLi_2H_{17}$ that is thermodynamically favorable at 200 GPa. The electronic band structures and DOS are given in Supplementary Fig. 3a. Phonon dispersions show no image frequencies at 200 GPa (Supplementary Fig. 3b), indicating that $CeLi_2H_{17}$ is dynamically stable. The EPC constant λ is estimated to be 0.45 and the predicted $T_c^E$ is only 14 K. This rather low value is attributed to the dominance of electronic DOS at the Fermi level from the *f*-electrons of Ce, which is unfavorable to coupling with the vibration of the H atoms, leading to the low λ and $T_c^E$ results.

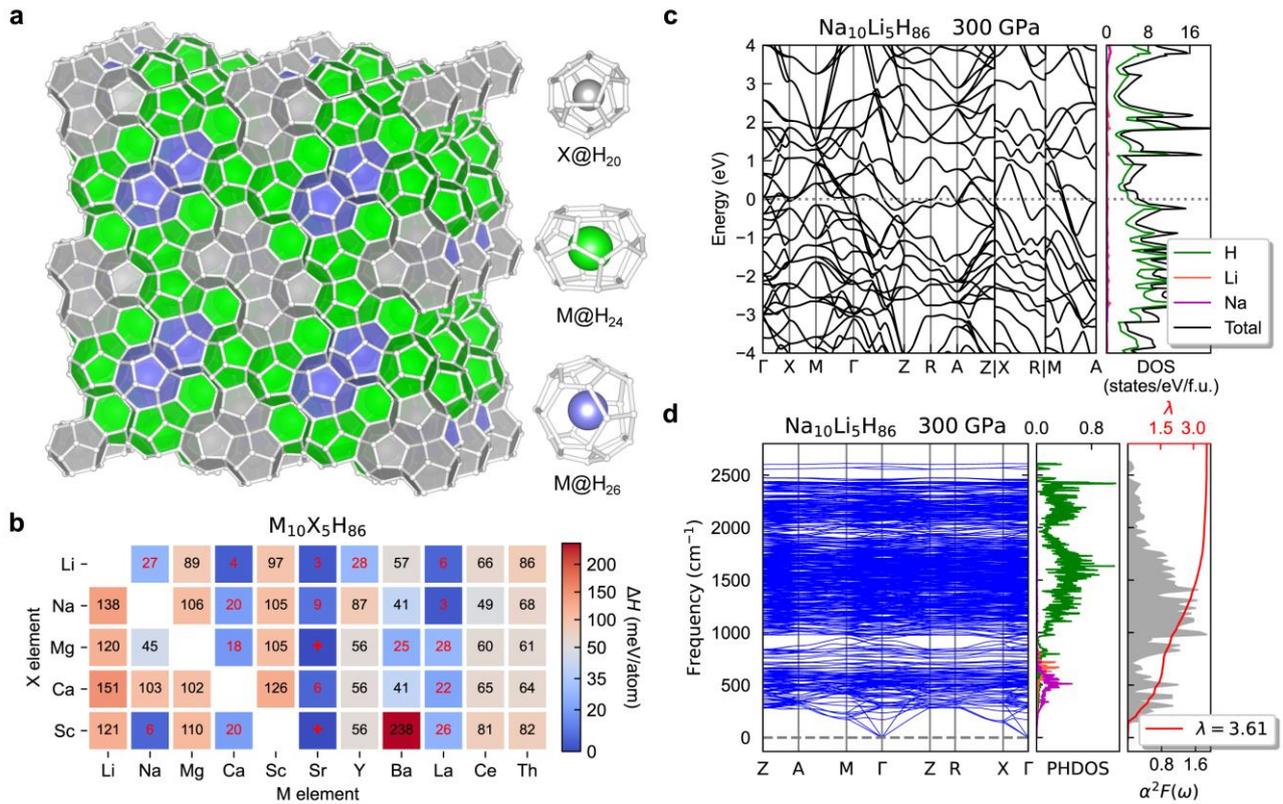

**Fig. 4 | Crystal structure, thermodynamic stability and electronic structure of type III clathrate hydrides.**
**a.** Structure and cage units of type III ternary hydrides $M_{10}X_5H_{86}$. **b.** Heatmap for thermodynamic stability of $M_{10}X_5H_{86}$ at 200 GPa. Colors/numbers in squares indicate Δ$H$. Metastable hydrides (Δ$H$<35 meV/atom) are marked by red numbers, while red crosses indicate thermodynamically stable hydrides. **c.** Electronic band structure (left) and density of states (right) of $Na_{10}Li_5H_{86}$ at 300 GPa. **d.** Phonon dispersion curves (left), phonon density of states (middle) and Eliashberg spectral function (right) of $Na_{10}Li_5H_{86}$ at 300 GPa.

**Type III clathrate hydrides**

Type III clathrate hydride $M_{10}X_5H_{86}$ in space group symmetry $P4_2/mnm$ comprises ten $H_{20}$ [$4^05^{12}6^0$] cages, sixteen $H_{24}$ [$4^05^{12}6^2$] cages and four $H_{26}$ [$4^05^{12}6^3$] cages, as shown in Fig. 4a. These ternary hydrides harbor metal atom X (X = Li, Na, Mg, Ca, Sc) with smaller radii in the $H_{20}$ cage, while larger sized metal atom M occupies the $H_{24}$ and $H_{26}$ cages. There are 202 atoms in the unit cell and the H/metal ratio is 5.73. This structure is completely new among hydrides, and previous structure search efforts have been unable to handle complex hydrides with such a large number of atoms in the unit cell. Thermodynamic stability of these hydrides is summarized in Fig. 4b, showing that $Sr_{10}Mg_5H_{86}$ and $Sr_{10}Sc_5H_{86}$ are thermodynamically stable at 200 GPa. Electronic band structures and DOS were given in Supplementary Fig. 4a-b. Since type III clathrate hydrides host a very large number of atoms in the unit cell, direct calculation of $T_c$ is unfeasible computationally. Alternatively, the modified AE model was used and produced $T_c^{AE}$ for $Sr_{10}Mg_5H_{86}$ and $Sr_{10}Sc_5H_{86}$ to be 73 K and 122 K, respectively (Supplementary Table 1). $Ca_{10}Li_5H_{86}$ is above the convex hull by 4 meV/atom at 200 GPa and becomes thermodynamically stable at 250 GPa (see Supplementary Fig. 5a for the convex hull of Li-Ca-H system at 250 GPa). Calculated $T_c^{AD}$ of $Ca_{10}Li_5H_{86}$ is 178 K at 200 GPa and drops to 168 K at 250 GPa. We carefully examined $Na_{10}Li_5H_{86}$ since the modified AE model gives it a very high $T_c^{AE}$ of 306 K. We constructed the convex hull of the Li-Na-H system at 300 GPa, and $Na_{10}Li_5H_{86}$ falls onto the hull, indicating its thermodynamic stability (Supplementary Fig. 6). It is noted that almost all of the $N(E_f)$ comes from electrons of the H atoms (Fig. 4c). The phonon dispersion curves indicate that $Na_{10}Li_5H_{86}$ is dynamically stable at 300 GPa (Fig. 4d). The $\lambda$ and $\omega_{log}$ were calculated to be 3.61 and 948 K, respectively. The vibrations dominated by Na and Li atoms with low frequencies contributed 40 % to $\lambda$ and the rest is contributed by the vibrations of H atoms with high frequencies of 1000-2600 cm$^{-1}$. The $T_c^E$ is calculated to be 325-341 K with $\mu^*$=0.10-0.13, suggesting that type III clathrate hydride $Na_{10}Li_5H_{86}$ is a promising room-temperature superconductor.

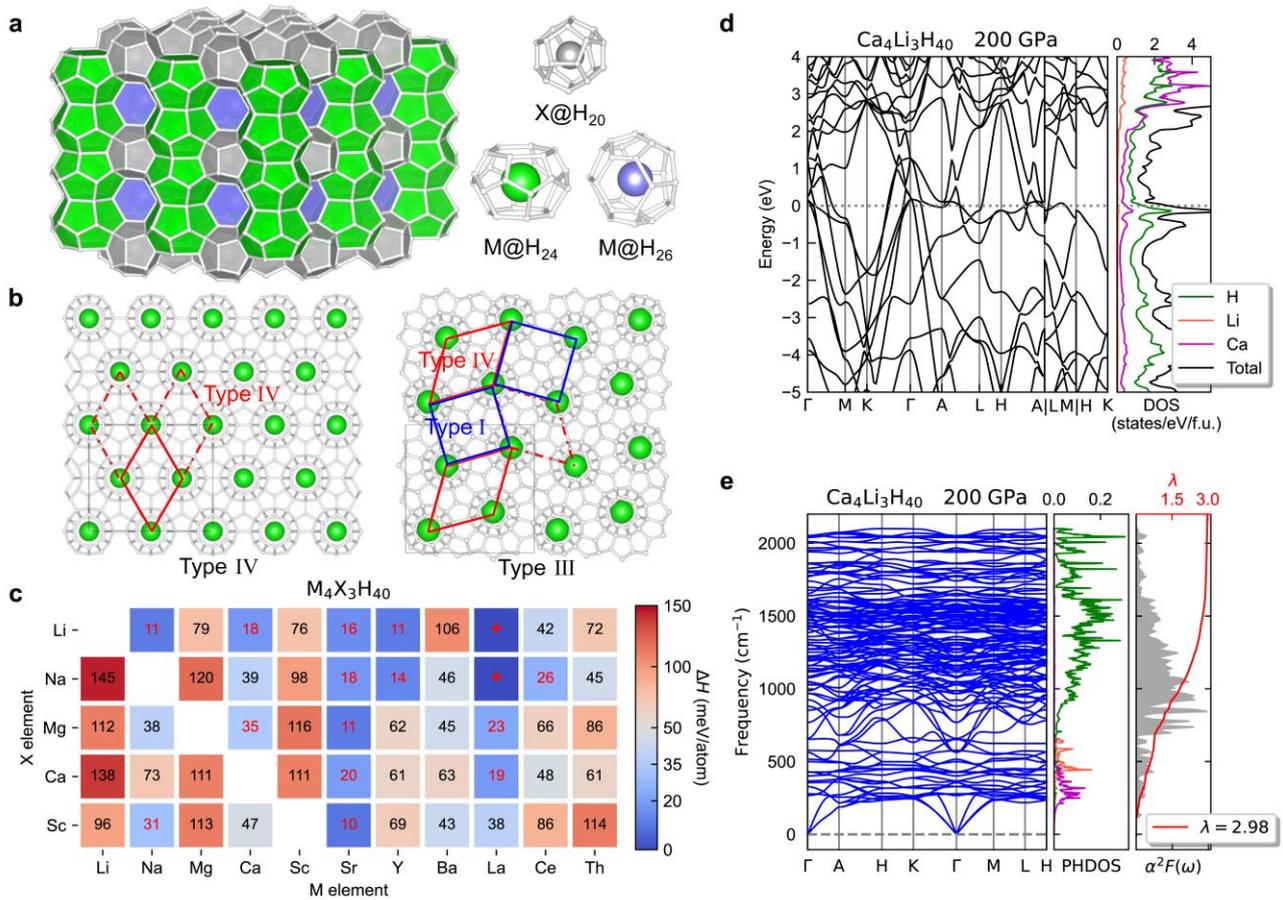

**Fig. 5 | Crystal structure, thermodynamic stability and electronic structure of type IV clathrate hydrides.** **a.** Structure and cage units of type IV ternary hydrides $M_4X_3H_{40}$. **b.** The supercell of type IV compared to that of type III structure. The latter can be viewed as a combination of type I (blue boxes) and type IV (red boxes). **c.** Heatmap of thermodynamic stability of $M_4X_3H_{40}$ at 200 GPa. Colors/numbers in squares indicate $\Delta H$. Metastable hydrides ($\Delta H$ <35 meV/atom) are marked by red numbers, while red crosses indicate thermodynamically stable hydrides. **d.** Electronic band structure (left) and density of states (right) of $Ca_4Li_3H_{40}$ at 200 GPa. **e.** Phonon dispersion curves (left), phonon density of state (middle) and Eliashberg spectral function (right) of $Ca_4Li_3H_{40}$ at 200 GPa.

**Type IV clathrate hydrides**

Type IV clathrate hydrides $M_4X_3H_{40}$ in $P6/mmm$ space group symmetry consists of three $H_{20}$ [$4^05^{12}6^0$] cages, two $H_{24}$ [$4^05^{12}6^2$] cages and two $H_{26}$ [$4^05^{12}6^3$] cages (Fig. 5a). These are the same cage units as in type III hydrides, but are arranged differently here. The metal atoms in type IV clathrate hydrides are arranged as in $P6/mmm$-$Zr_4Al_3$ and the H/metal ratio is 5.71. To better compare the structures of type IV and III clathrate hydrides, we constructed a supercell of type IV clathrate hydrides with a tetragonal structure that is similar to type III structure, as shown in Fig. 5b. The hexagonal unit cells of type IV clathrate structure also exist in type III structure, and they are arranged in an "angle-to-angle" pattern in type III structure but "edge-to-edge" in type IV. In fact, type III clathrate comprises type I and type IV, as shown in the right panel in Fig. 5b.

The heatmap for thermodynamic stability for type IV clathrate hydrides is given in Fig. 5c, and $La_4Li_3H_{40}$ and $La_4Na_3H_{40}$ are thermodynamically viable at 200 GPa. The convex hull containing ternary hydrides from previous studies confirm thermodynamic stability of $La_4Li_3H_{40}$ at 200 GPa (Supplementary Fig. 7). The electronic band structures and DOS of $La_4Li_3H_{40}$ are shown in Supplementary Fig. 8a. Only a few bands cross the Fermi level resulting in a small $N(E_f)$. The calculated $\lambda$ is 1.01, producing $T_c^E$ with $\mu^*$=0.10 of 93 K (Supplementary Fig. 8b). $La_4Na_3H_{40}$ exhibits dynamic instability below 300 GPa and is not further examined. We also studied metastable hydride $Ca_4Li_3H_{40}$ that is above the convex hull by mere 18 meV/atom at 200 GPa. Its $T_c^{AE}$ is 294 K, suggesting potential for being a room-temperature superconductor. A von Hove singularity near the Fermi level from electrons of H atoms dominates $N(E_f)$ (Fig. 5d). Phonon dispersion curves in Fig. 5e show no imaginary frequency suggesting that $Ca_4Li_3H_{40}$ is dynamically stable at 200 GPa. The EPC constant $\lambda$ is 2.98, with main contribution from vibrations of H atoms in the frequency range of 750-1600 cm$^{-1}$. The $T_c^E$ with $\mu^*$=0.1 is 293 K, in excellent agreement with $T_c^{AE}$ of 294 K, validating the reliability of the AE model.

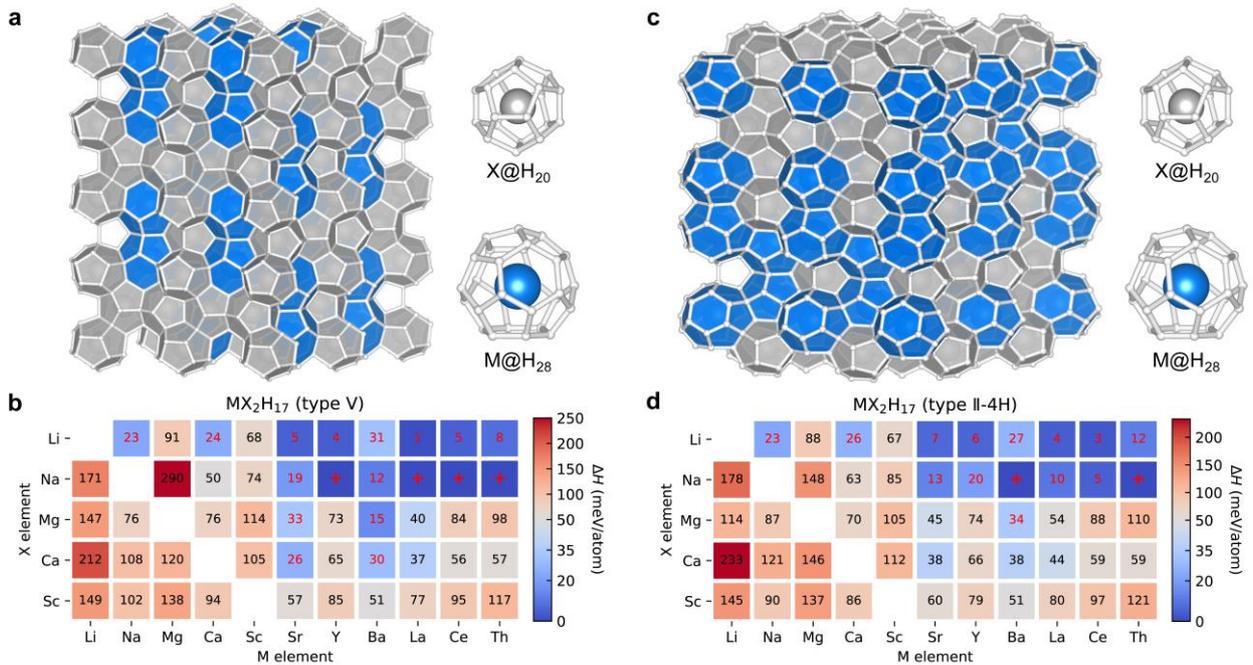

**Fig. 6 | Crystal structure and thermodynamic stability of type V and II-4H clathrate hydrides.** Structure and cage units of ternary hydrides (**a**) $MX_2H_{17}$ with type V structure and (**c**) $MX_2H_{17}$ with type II-4H structure. Heatmap of thermodynamic stability of (**b**) type V and (**d**) type II-4H clathrate hydrides. Colors/numbers in squares indicate $\Delta H$. The metastable hydrides ($\Delta H$<35 meV/atom) are marked by red numbers, while red crosses indicate thermodynamically stable hydrides.

**Type V and type II-4H clathrate hydrides**

Type V and II-4H clathrate hydrides share the same stoichiometry of $MX_2H_{17}$ with type II clathrate hydrides. The unit cell of type V structure contains four formula units and type II-4H structure contains eight. The three structures all consist of $H_{20}$ $[4^05^{12}6^0]$ cages and $H_{28}$ $[4^05^{12}6^4]$ cages but with different arrangements. The metal atoms in type V and II-4H structures are arranged as in $P6_3/mmc$-$MgZn_2$ and $P6_3/mmc$-$MgNi_2$, respectively (Fig. 6a and 6c). Thermodynamic stability status is summarized in Fig. 6b and 6d. In general, type V and type II-4H clathrate hydrides have similar formation enthalpy. We find $YNa_2H_{17}$, $LaNa_2H_{17}$ and $CeNa_2H_{17}$ in type V structure and $BaNa_2H_{17}$ in type II-4H structure are thermodynamically viable at 200 GPa. Their electronic band structures and DOS are given in Supplementary Fig. 9. For $YNa_2H_{17}$, there are few bands crossing the Fermi level, leading to very small $N(E_f)$ that is unfavorable to superconductivity. The EPC constant $\lambda$ and $T_c^{AD}$ of $YNa_2H_{17}$ at 200 GPa are 0.57 and 5 K, respectively. Small $N(E_f)$ also occurs in type V $LaNa_2H_{17}$ and type II-4H phase $BaNa_2H_{17}$, with the calculated $T_c^{AD}$ of 17 K and 98 K, respectively. The $f$ electron of Ce atoms dominates the $N(E_f)$ of type V $CeNa_2H_{17}$ with $T_c$ of 29 K at 200 GPa.

Based on high $T_c^{AE}$ values predicted by the modified AE model, we focus on $NaLi_2H_{17}$, $CaLi_2H_{17}$ and $SrLi_2H_{17}$ in type V clathrate structure. $NaLi_2H_{17}$ (type V) is located 23 meV/atom above the convex hull at 200 GPa. At 300 GPa, this value decreases to only 3 meV/atom, raising the prospect of experimental synthesis. Type II, type V and type II-4H structures are competing phases with the formation enthalpy difference within 2 meV/atom at 200 GPa (Supplementary Fig. 10). $NaLi_2H_{17}$ (type V) possesses a von Hove singularity at the Fermi level and almost all the $N(E_f)$ are contributed by the electrons of H atom (Fig. 7a). It is dynamically stable down to 220 GPa, where $T_c^E$ is 336 K (Fig. 7b). For comparison, $T_c^E$ of type II-4H phase $NaLi_2H_{17}$ at 200 GPa is 326 K (Supplementary Fig. 11 and Table 2). For $CaLi_2H_{17}$ (type V), it is thermodynamically stable above 320 GPa (Supplementary Fig. 5b). The electronic band structures and DOS at 250 GPa (Fig. 7c) show several flat bands close to the Fermi level along the high symmetry path of A to L and H to A, resulting in a large $N(E_f)$. The phonon dispersion curves suggest its dynamic stability at 250 GPa (Fig. 7d). Its $\lambda$ and $\omega_{log}$ are 3.03 and 1187 K, respectively. The $\lambda$ is mainly contributed by the vibrations of atomic H with frequency of 800-2300 cm$^{-1}$, and the phonon modes with frequency below 800 cm$^{-1}$ contribute 37 % to the $\lambda$. $T_c^E$ is 326 K with $\mu^*=0.1$. For $SrLi_2H_{17}$ (type V), it is thermodynamically stable at 300 GPa (Supplementary Fig. 12a). The electronic band structures and DOS in Fig. 7e are very similar to those of $CaLi_2H_{17}$ since elements Sr and Ca have the same number of valence electrons and similar properties. There are also flat bands near the Fermi level and the large values of $N(E_f)$ are dominated by H atoms electrons. Phonon modes with frequencies above and below 800 cm$^{-1}$ contribute almost equally to $\lambda$ at 53% and 47%, respectively (Fig. 7f). The $\lambda$ and $T_c^E$ are 3.27 and 309 K, respectively.

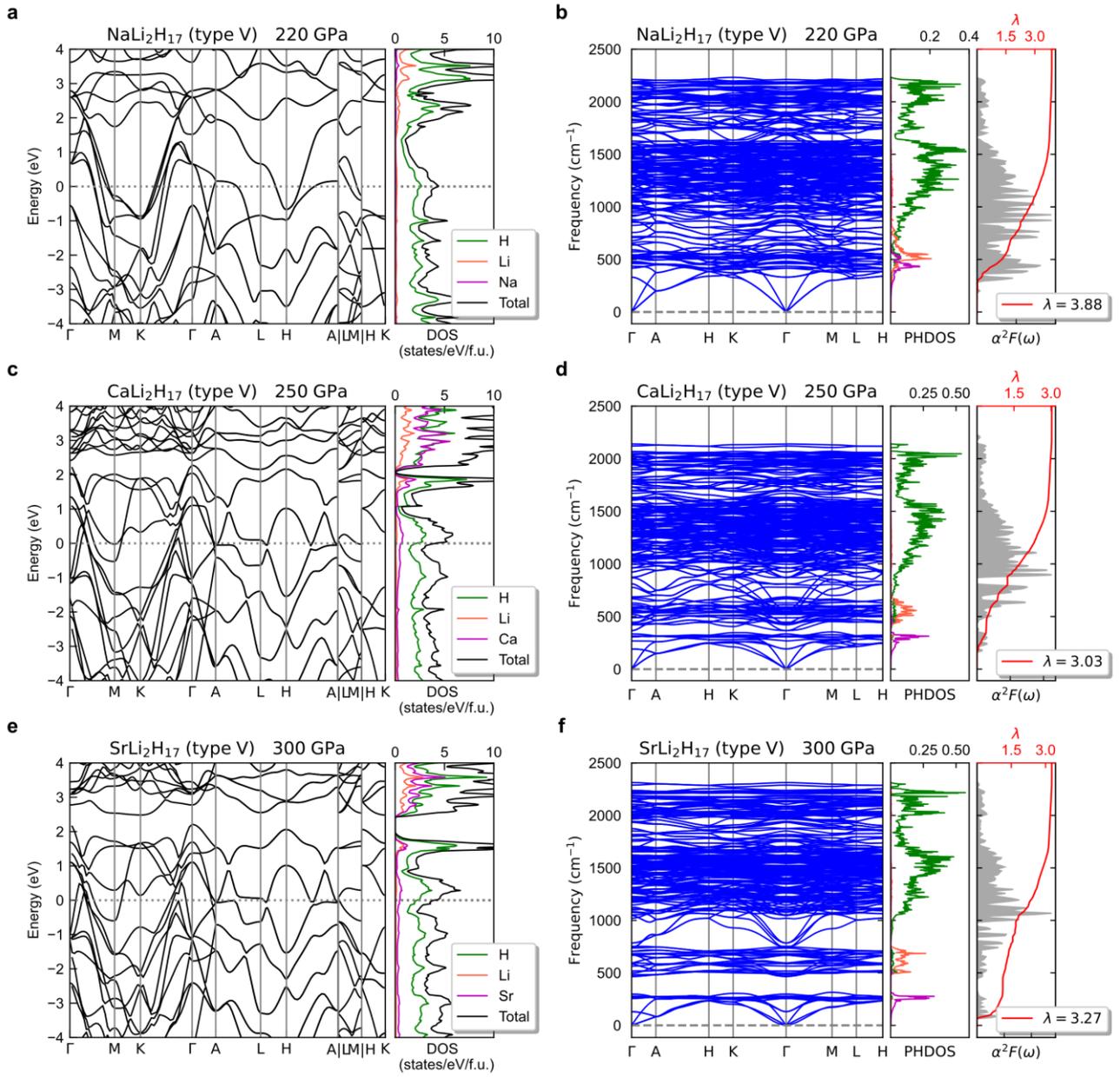

**Fig. 7 | Electronic band structures and phonon dispersion curves of type V clathrate hydrides.** Electronic band structures (left) and density of states (right) of (**a**) NaLi$_2$H$_{17}$ at 220 GPa, (**c**) CaLi$_2$H$_{17}$ at 250 GPa, (**e**) SrLi$_2$H$_{17}$ at 300 GPa. Phonon dispersion curves (left), phonon density of states (middle) and Eliashberg spectral function (right) of (**b**) NaLi$_2$H$_{17}$ at 220 GPa, (**d**) CaLi$_2$H$_{17}$ at 250 GPa, (**f**) SrLi$_2$H$_{17}$ at 300 GPa.

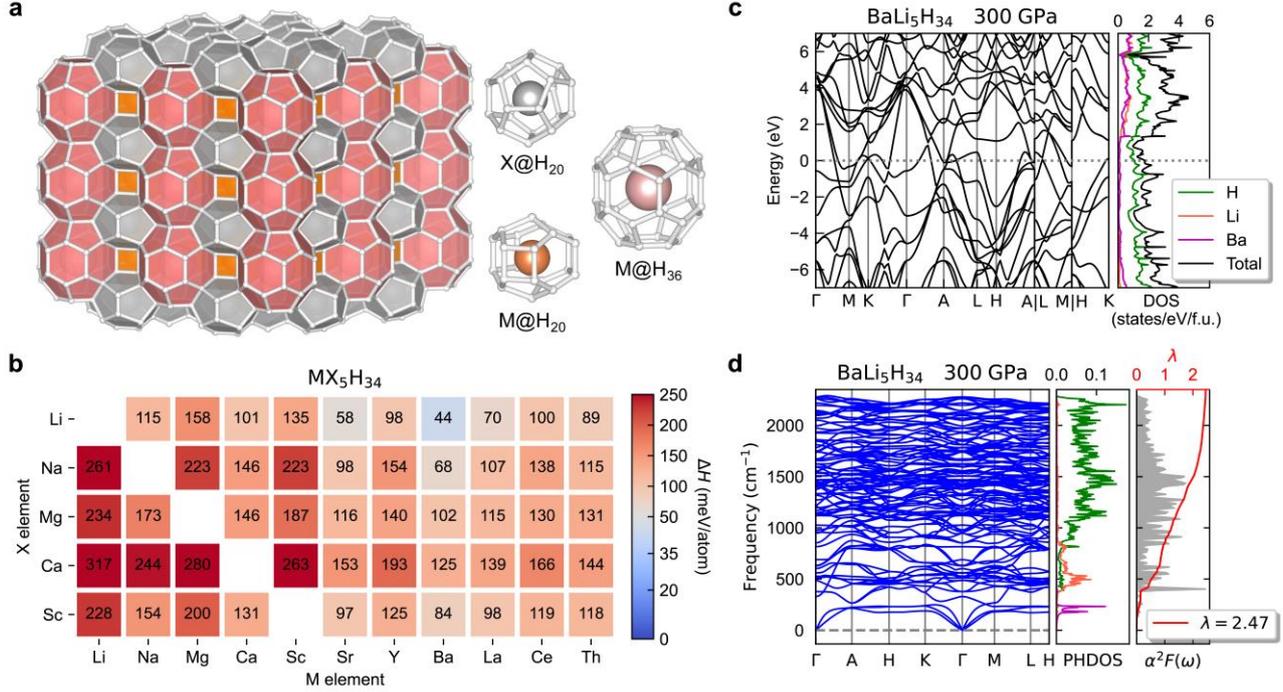

**Fig. 8 | Crystal structure, thermodynamic stability and electronic structure of type H clathrate hydrides.**
**a.** Structure and cage units of type H ternary hydrides $MX_5H_{34}$. **b.** Heatmap of thermodynamic stability of $MX_5H_{34}$ at 200 GPa. Colors/numbers in squares indicate $\Delta H$. **c.** Electronic band structures (left) and density of states (right) of $BaLi_5H_{34}$ at 300 GPa. **d.** Phonon dispersion curves (left), phonon density of states (middle) and Eliashberg spectral function (right) of $BaLi_5H_{34}$ at 300 GPa.

**Type H clathrate hydrides**

Type H structure $MX_5H_{34}$ in space group symmetry *P6/mmm* consists of three $H_{20}$ [$4^05^{12}6^0$] cages, three $H_{20}$ [$4^35^66^3$] cages and one $H_{36}$ cage [$4^05^{12}6^8$] (Fig. 8a), with the same metal atoms in the two types of $H_{20}$ cages and the H/metal ratio is 5.71. Thermodynamic stability assessment is shown in Fig. 8b, and there is no thermodynamically stable hydride at 200 GPa. The $H_{36}$ cage is one of the largest cages in the clathrate hydrides and it seems that the metal atoms are not large enough to support the $H_{36}$ cages. A previous study reported $H_{36}$ cage structure in $CeH_{18}$[46], but higher pressure of 350 GPa is needed to stabilize this structure. It is expected that higher pressure is also required to stabilize type H clathrate hydrides studied in this work. Meanwhile, $BaLi_5H_{34}$ containing metal atoms of larger radii occupying $H_{36}$ cages possesses relatively good thermodynamic stability, deviating from the convex hull by only 23 meV/atom at 300 GPa, where it is dynamically stable. Its electronic band structures and DOS are given in Fig. 8c. The $N(E_f)$ of $BaLi_5H_{34}$ is contributed mainly by the electrons of H atoms, which is favorable to superior superconductivity. The EPC constant $\lambda$ is 2.47, most of which is contributed by the vibrations of H atoms with the frequencies of 1000-2000 cm$^{-1}$ (Fig. 8d), producing $T_c^E$ of 295 K at 300 GPa with $\mu^*$=0.10.

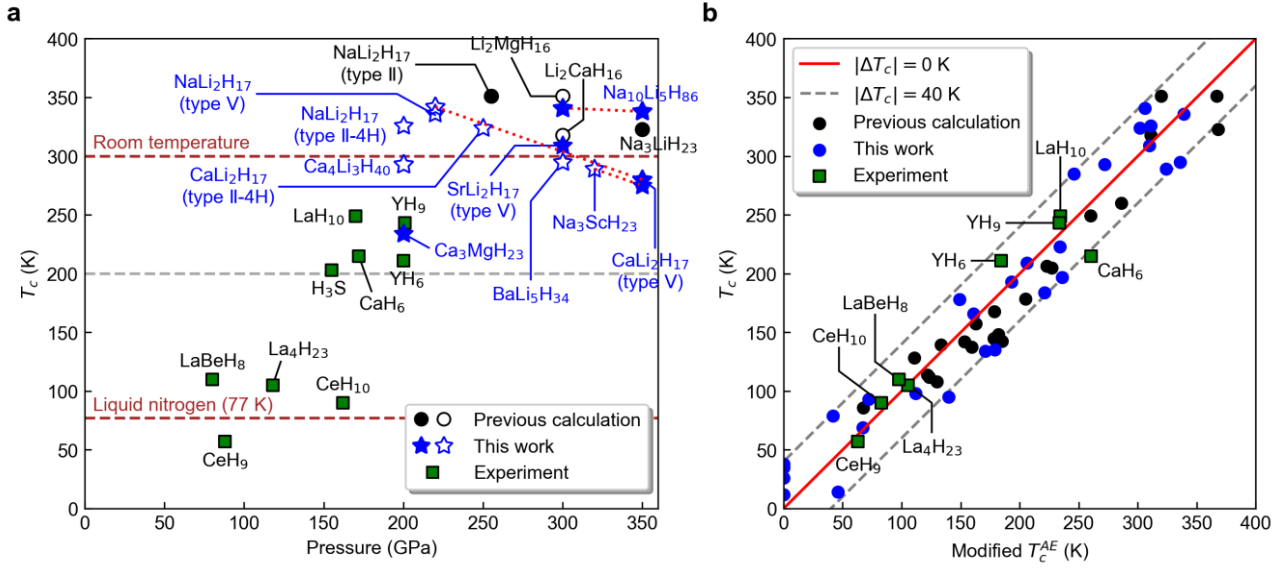

**Fig. 9 | Summary of experimental and theoretical $T_c$ values of prominent hydrides. a.** Scatter plot of $T_c$ versus pressure for hydrides that have been experimentally synthesized (green squares), theoretically predicted in previous works (black circles) and in this work (blue filled and unfilled stars for thermodynamically stable and metastable hydrides, respectively). **b.** Scatter plot of experimentally measured (green squares) and theoretically calculated (circles) $T_c$ versus $T_c$ estimated by the modified AE model. The red and the grey dashed lines signify that the errors for the $T_c$ estimated by the modified AE model is 0 K and 40 K, respectively.

## DISCUSSION

**Superconductivity and the modified AE mode**

The high-$T_c$ clathrate hydrides uncovered in this study are summarized in Fig. 9a, compared with those of experimentally measured and predicted by previous studies. Ten most prominent and promising high-$T_c$ ternary clathrate hydrides discovered in this work are $Na_3ScH_{23}$ (type I), $Ca_3MgH_{23}$ (type I), $Na_{10}Li_5H_{86}$ (type III), $Ca_4Li_3H_{40}$ (type IV), $NaLi_2H_{17}$ (type V, type II-4H), $CaLi_2H_{17}$ (type V, type II-4H), $SrLi_2H_{17}$ (type V), and $BaLi_5H_{34}$ (type H). Type I clathrate hydride $Na_3ScH_{23}$ and Type III clathrate hydride $Na_{10}Li_5H_{86}$ are both thermodynamically and dynamically stable above 350 GPa and 300 GPa, respectively, with $T_c^E$ of 275 K and 341 K. Type V clathrate hydride $CaLi_2H_{17}$ is thermodynamically stable at 320 GPa and is dynamically stable down to 220 GPa with $T_c^E$ of 342 K. $SrLi_2H_{17}$ in type V clathrate structure possesses thermodynamic stability at 300 GPa with $T_c^E$ of 309 K. More detailed data on superconductivity of the clathrate structures studied in this work are given in Supplementary Table 2. A comprehensive comparison of $T_c$ values of these clathrate hydrides allows a systematic examination of the validity of the modified AE model. A scatter plot of calculated and measured $T_c$ values compared with $T_c^{AE}$ estimated by the modified AE model are shown in Fig. 9b. The MAE is 24.5 K and most of the prediction errors are limited within 40 K, suggesting that the modified AE model can predict the $T_c$ of clathrate hydrides with reasonable accuracy at much reduced computational costs.

**Thermodynamic stability and the match factor**

The HTP data provide a good platform to summarize the general rules for thermodynamic stability of clathrate hydrides. Geometric parameters are important factors in the high-throughput screening for materials like perovskites[47] and MXenes[48]. For clathrate hydrides, the radii of the metal atoms should match the size of the H cages. For example, Ca atoms can well support $H_{24}$ cages, whereas hydrides with lanthanide metals with larger radii such as La and Ce are more inclined to form clathrate structures containing larger cages such as $H_{32}$ and $H_{29}$ cages. Here we quantify a match factor ($\rho$) between the sizes of metal atoms and cages and analyze its relationship with thermodynamic stability. We simplify the metal atoms as hard spheres, and the volume of metal atom $M$ is given by

$$v_M = \frac{4}{3}\pi r_m^3, \qquad (3)$$

where $r_m$ denotes the radii of metal atoms $M^{49,50}$. We then define a "standard clathrate structure" for each type, which is structurally optimized by software RG2[51], which supports structural geometry optimization for only the H cages based on given bond lengths and angles, avoiding the influence of metal atoms on cage size. We take the geometrically optimized bond length of 1.15 Å and bond angle of 106.5° and obtain unbiased "standard clathrate structures". We define the matching relation between the absolute volume of metal atoms and the cages by the formula

$$\rho_a = \frac{\sum_M \left(1 - \frac{|v_M - V_M|}{V_M}\right)}{n}, (M = 1,2,...,n), \qquad (4)$$

where $V_M$ denotes the volume of the H cage in the "standard clathrate structures" and $n$ is the number of the metal atoms in the unit cell. In addition, we introduce $\rho_r$ to measure the matching relationship between the volume ratio of different cages and that of the corresponding metal atoms to give a scaling tolerance to the "standard clathrate structure". Taking the example of a ternary hydride containing two different cages with metal atoms X and Y, the $\rho_r$ is defined by the formula

$$\rho_r = 1 - \frac{\left|\frac{v_X}{v_Y} - \frac{V_X}{V_Y}\right|}{\frac{V_X}{V_Y}}. \qquad (5)$$

Finally, $\rho$ is defined as

$$\rho = \frac{3}{4}\rho_a + \frac{1}{4}\rho_r. \qquad (6)$$

The calculated results of $\rho$ for different types of clathrate hydrides are summarized in Fig. 10. For ease of comparison with thermodynamic stability, the figure shows the data of $1 - \rho$. There is a clear and strong correlation between clathrate hydrides with $\rho > 0.75$ and metastability with $\Delta H < 35$ meV/atom. Specifically, the percentages of stable or metastable clathrate hydrides agreeing with large $\rho$ value defined above are 64%, 64%, 72%, 69%, 56%, 67% and 100% for type I, II, III, IV, V, II-4H and H clathrate hydrides, respectively. At the same time, the number of hydrides still need to be calculated after screened by the match factor is drastically reduced to 15% (type I), 12% (type II), 20% (type III), 37% (type IV), 20% (type V), 20% (type II-4H) and 8% (type H). This finding suggests that $\rho$ can well catch thermodynamic stability of clathrate hydrides with little computational costs. However, since the lanthanides (e.g. La and Ce) have similar radii, so the match factor cannot distinguish between them. As a result, thermodynamic stability predicted by the match factors of clathrate hydrides containing Ce and La atoms are similar, as shown in the bottom of Fig.10a, 10c and 10d. Overall, the match factor can help to screen for thermodynamically stable clathrate hydrides with little computational costs and facilitate high-throughput calculations of other clathrate hydrides with diverse and complex structures with larger unit cells.

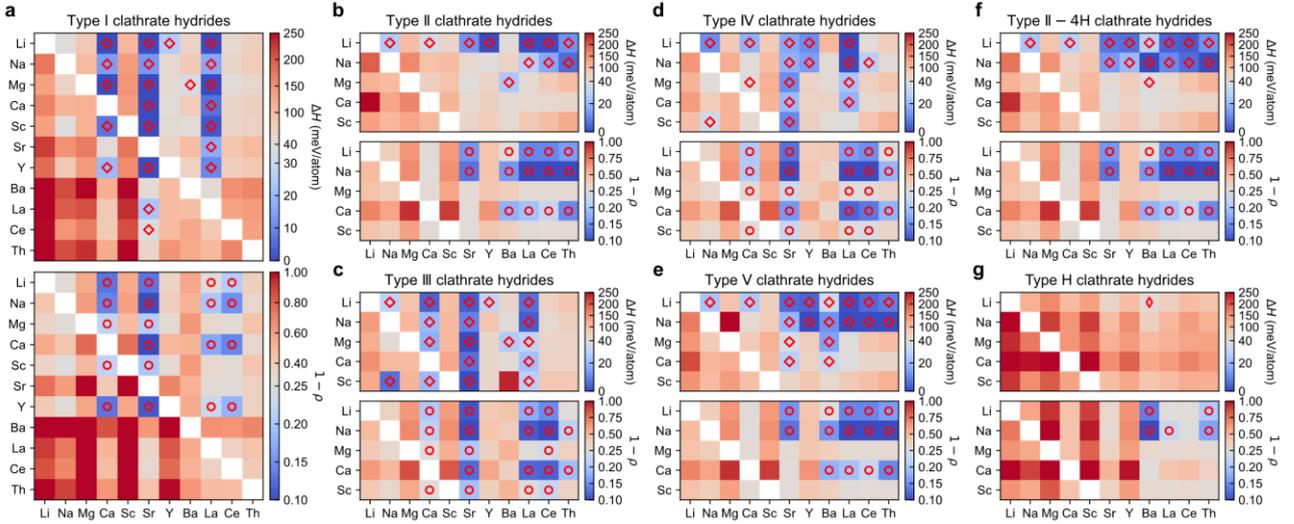

**Fig. 10 | Thermodynamic stability and match factors.** Heatmap of thermodynamic stability and the match factor for **a**. type I, **b**. type II, **c**. type III, **d**. type IV, **e**. type V, **f**. type II-4H **g**. type H clathrate hydrides. Diamond markers indicate metastable hydrides with $\Delta H < 35$ meV/atom and circles indicate clathrate hydrides with match factor $\rho > 0.75$. The thin diamond marker in **e**. indicates $BaLi_5H_{34}$ with minimum $\Delta H$ of 44 meV in type H clathrate hydrides.

## CONCLUSION

In summary, we have found a series of ternary multi-cage clathrate hydrides with non-integer H/metal ratios by a design principle that emulates the clathrate structures of hydrates and group 14 element frameworks. The central idea is a high-throughput evaluation for thermodynamic stability using a self-built hydride database, jointly with a rapid assessment of superconducting critical temperature $T_c$ by an efficient modified AE model. This process leads to the discovery of an eclectic collection of nineteen complex ternary hydrides in distinct categories of structures that are thermodynamically stable at readily accessible pressure of 200 GPa, i.e. $Ca_3(Li, Mg)H_{23}$, $Sr_3(Li, Mg, Ca, Sc, Y)H_{23}$, $La_3(Li, Mg)H_{23}$ in type I structure, $CeLi_2H_{17}$ in type II structure, $Sc_{10}Na_5H_{86}$, $Sr_{10}(Na, Sc)_5H_{86}$ in type III structure, $La_4Li_3H_{40}$ in type IV structure, $(Y, La, Ce)Na_2H_{17}$ in type V structure and $BaNa_2H_{17}$ in type II-4H structure. Additional metastable hydrides close to the convex hull with $\Delta H$ less than 35 meV/atom also have been identified. Prominent among the thermodynamically stable clathrate hydrides,

Na$_3$ScH$_{23}$ (type I), CaLi$_2$H$_{17}$ (type V), Na$_{10}$Li$_5$H$_{86}$ (type III) and SrLi$_2$H$_{17}$ (type V) host $T_c$ around 300 K at 350 GPa, 320 GPa, 300 GPa and 300 GPa, respectively, making them promising room temperature superconductors that are possible to be experimentally synthesized. Besides, metastable hydrides Ca$_4$Li$_3$H$_{40}$ (type IV), CaLi$_2$H$_{17}$ (type II-4H), NaLi$_2$H$_{17}$ (type V and II-4H) and BaLi$_5$H$_{34}$ (type H) also show near-room-temperature $T_c$. Moreover, we introduce a match factor based on the relative sizes of the metal atoms and hydrogen cages that is well correlated with thermodynamic stability, which now can be assessed highly efficiently with very low computational costs. This approach establishes a new paradigm to accelerate discovery of thermodynamically viable room-temperature superconductors among complex ternary or higher-order clathrate hydrides with a very large number of atoms in their unit cells, which are beyond the reach of prevailing structure search and property calculation methods. The present work introduces an innovative way to expand the accessible realm of complex clathrate hydrides that harbor rich chemical bonding features and novel physical properties.

## METHODS

### Density functional theory (DFT) calculations

The DFT calculations for the database were performed using VASP code[52]. The PAW pseudopotentials with PBE exchange-correlation functional under the generalized gradient approximation (GGA) were selected[53]. For both efficiency and precision, the cutoff energy and spacing of k points mesh were selected as 600 eV and $2\pi \times 0.03$ Å$^{-1}$. We also employed the VASP code to calculate the electronic band structures, density of states (DOS) and electron localization function (ELF). The pseudopotentials and parameters were selected in the same way as those used in building the database whereas the cutoff energy was increased to 800 eV.

### Electron-phonon coupling and superconducting $T_c$ calculations

The Quantum Espresso (QE) package[45] was employed to calculate the phonon dispersion and electron-phonon coupling constant. Norm-conserving pseudopotentials[54] were employed; for calculations involving elements with f-electrons such as La and Ce, PAW pseudopotentials[55] considering f-electrons were selected. The kinetic energy cutoff was 100 Ry for wavefunctions and 800 Ry for charge density and potential. The information on the accuracy of q-mesh and k-mesh selected for each type of clathrate hydrides are given in Supplementary Table 3. The superconducting transition temperature ($T_c$) was estimated using the Allen–Dynes modified McMillan (A-D) equation[56,57] and self-consistent solution of the Eliashberg function[58].

### Structure search

The AIRSS software[59] was employed for structure search combined with the DFT software CASTEP[60]. Ultrasoft pseudopotentials were employed and the energy cutoff and k-points spacing were selected as 400 eV and $2\pi \times 0.07$ Å$^{-1}$, respectively. For calculations to determine the convex hull, VASP codes were used, with the use of the same parameters as in constructing the database.

# References

1   Ashcroft, N. W. Metallic Hydrogen: A High-Temperature Superconductor? *Phys. Rev. Lett.* **21**, 1748-1749 (1968).
2   Eremets, M. I., Drozdov, A. P., Kong, P. P. & Wang, H. Semimetallic molecular hydrogen at pressure above 350 GPa. *Nat. Phys.* **15**, 1246-1249 (2019).
3   Monacelli, L., Casula, M., Nakano, K., Sorella, S. & Mauri, F. Quantum phase diagram of high-pressure hydrogen. *Nat. Phys.* **19**, 845-850 (2023).
4   Loubeyre, P., Occelli, F. & Dumas, P. Synchrotron infrared spectroscopic evidence of the probable transition to metal hydrogen. *Nature* **577**, 631-635 (2020).
5   Ashcroft, N. W. Hydrogen dominant metallic alloys: high temperature superconductors? *Phys. Rev. Lett.* **92**, 187002 (2004).
6   Duan, D. *et al.* Pressure-induced metallization of dense $(H_2S)_2H_2$ with high-$T_c$ superconductivity. *Sci. Rep.* **4**, 6968 (2014).
7   Drozdov, A. P., Eremets, M. I., Troyan, I. A., Ksenofontov, V. & Shylin, S. I. Conventional superconductivity at 203 kelvin at high pressures in the sulfur hydride system. *Nature* **525**, 73-76 (2015).
8   Einaga, M. *et al.* Crystal Structure of the Superconducting Phase of Sulfur Hydride. *Nat. Phys.* **12**, 835-838 (2016).
9   Liu, H., Naumov, II, Hoffmann, R., Ashcroft, N. W. & Hemley, R. J. Potential high-Tc superconducting lanthanum and yttrium hydrides at high pressure. *Proc. Natl. Acad. Sci. U.S.A.* **114**, 6990-6995 (2017).
10  Peng, F., Sun, Y., Pickard, C. J., Needs, R. J., Wu, Q. & Ma, Y. Hydrogen Clathrate Structures in Rare Earth Hydrides at High Pressures: Possible Route to Room-Temperature Superconductivity. *Phys. Rev. Lett.* **119**, 107001 (2017).
11  Xie, H. *et al.* Hydrogen Pentagraphenelike Structure Stabilized by Hafnium: A High-Temperature Conventional Superconductor. *Phys. Rev. Lett.* **125** (2020).
12  Wang, H., Tse, J. S., Tanaka, K., Iitaka, T. & Ma, Y. Superconductive sodalite-like clathrate calcium hydride at high pressures. *Proc. Natl. Acad. Sci. U.S.A.* **109**, 6463-6466 (2012).
13  Ma, L. *et al.* High-Temperature Superconducting Phase in Clathrate Calcium Hydride $CaH_6$ up to 215 K at a Pressure of 172 GPa. *Phys. Rev. Lett.* **128**, 167001 (2022).
14  Li, Z. *et al.* Superconductivity above 200 K discovered in superhydrides of calcium. *Nat. Commun.* **13**, 2863 (2022).
15  Drozdov, A. P. *et al.* Superconductivity at 250 K in lanthanum hydride under high pressures. *Nature* **569**, 528-531 (2019).
16  Somayazulu, M. *et al.* Evidence for Superconductivity above 260 K in Lanthanum Superhydride at Megabar Pressures. *Phys. Rev. Lett.* **122**, 027001 (2019).
17  Chen, W. *et al.* High-Temperature Superconducting Phases in Cerium Superhydride with a $T_c$ up to 115 K below a Pressure of 1 Megabar. *Phys. Rev. Lett.* **127**, 117001 (2021).
18  Semenok, D. V. *et al.* Superconductivity at 161 K in thorium hydride $ThH_{10}$: Synthesis and properties. *Mater. Today* **33**, 36-44 (2020).
19  Salke, N. P. *et al.* Synthesis of clathrate cerium superhydride $CeH_9$ at 80-100 GPa with atomic hydrogen sublattice. *Nat. Commun.* **10**, 4453 (2019).
20  Snider, E. *et al.* Synthesis of Yttrium Superhydride Superconductor with a Transition Temperature up to 262 K by Catalytic Hydrogenation at High Pressures. *Phys. Rev. Lett.* **126**, 117003 (2021).
21  Kong, P. *et al.* Superconductivity up to 243 K in the yttrium-hydrogen system under high pressure. *Nat. Commun.* **12**, 5075 (2021).

22 Semenok, D. V. *et al.* Superconductivity at 253 K in lanthanum–yttrium ternary hydrides. *Mater. Today* **48**, 18-28 (2021).

23 Bi, J. *et al.* Giant enhancement of superconducting critical temperature in substitutional alloy (La,Ce)H$_9$. *Nat. Commun.* **13**, 5952 (2022).

24 Chen, W. *et al.* Enhancement of superconducting properties in the La-Ce-H system at moderate pressures. *Nat. Commun.* **14**, 2660 (2023).

25 Semenok, D. V. *et al.* Novel Strongly Correlated Europium Superhydrides. *J. Phys. Chem. Lett.* **12**, 32-40 (2021).

26 Pena-Alvarez, M. *et al.* Synthesis of Weaire-Phelan Barium Polyhydride. *J. Phys. Chem. Lett.* **12**, 4910-4916 (2021).

27 Cross, S. *et al.* High-temperature superconductivity in La$_4$H$_{23}$ below 100 GPa. *Phys. Rev. B* **109**, L020503 (2024).

28 Laniel, D. *et al.* High-pressure synthesis of seven lanthanum hydrides with a significant variability of hydrogen content. *Nat. Commun.* **13**, 6987 (2022).

29 Li, Z. *et al.* Superconductivity above 70 K observed in lutetium polyhydrides. *Sci. China Phys. Mech. Astron.* **66**, 267411 (2023).

30 Diverse high-pressure chemistry in Y-NH$_3$BH$_3$ and Y–paraffin oil systems. *Sci. Adv.* **10**, eadl5416 (2024).

31 Sun, Y., Wang, Y., Zhong, X., Xie, Y. & Liu, H. High-temperature superconducting ternary Li−R−H superhydrides at high pressures (R=Sc,Y,La). *Phys. Rev. B* **106**, 024519 (2022).

32 Sun, Y., Lv, J., Xie, Y., Liu, H. & Ma, Y. Route to a Superconducting Phase above Room Temperature in Electron-Doped Hydride Compounds under High Pressure. *Phys. Rev. Lett.* **123**, 097001 (2019).

33 An, D., Duan, D., Zhang, Z., Jiang, Q., Song, H. & Cui, T. Thermodynamically stable room temperature superconductors in Li-Na hydrides under high pressures. Preprint at (2024).

34 Predicted hot superconductivity in LaSc$_2$H$_{24}$ under pressure. *Proceedings of the National Academy of Sciences* **121**, e2401840121 (2024).

35 Sun, Y. & Miao, M. Assemble superhydrides with non-integer H/Metal ratios on metal templates. Preprint at https://doi.org:10.48550/arXiv.2303.05721 (2024).

36 Manakov, A. Y., Kosyakov, V. I. & Solodovnikov, S. F. in *Comprehensive Supramolecular Chemistry II Ch. 7* (Elsevier, Netherlands, 2017).

37 Karttunen, A. J., Fassler, T. F., Linnolahti, M. & Pakkanen, T. A. Structural principles of semiconducting Group 14 clathrate frameworks. *Inorg. Chem.* **50**, 1733-1742 (2011).

38 Ma, T., Zhang, Z., Wu, S., Duan, D. & Cui, T. FF7: a code package for high-throughput calculations and constructing materials database. In preparation (2024).

39 Zhang, Z. H. *et al.* Design Principles for High-Temperature Superconductors with a Hydrogen-Based Alloy Backbone at Moderate Pressure. *Phys. Rev. Lett.* **128**, 047001 (2022).

40 Lucrezi, R., Di Cataldo, S., von der Linden, W., Boeri, L. & Heil, C. In-silico synthesis of lowest-pressure high-$T_c$ ternary superhydrides. *Npj Comput. Mater* **8**, 119 (2022).

41 Di Cataldo, S., von der Linden, W. & Boeri, L. First-principles search of hot superconductivity in La-X-H ternary hydrides. *NPJ Comput. Mater* **8**, 2 (2022).

42 Du, J. *et al.* First-principles study of high-pressure structural phase transition and superconductivity of YBeH$_8$. *J. Chem. Phys* **160**, 094116 (2024).

43 Sun, Y., Sun, S., Zhong, X. & Liu, H. Prediction for high superconducting ternary hydrides below megabar pressure. *J. Phys. Condens. Matter* **34**, 505404 (2022).

44 Ma, T. *et al.* High-throughput calculation for superconductivity of sodalite-like clathrate ternary hydrides MXH$_{12}$ at high pressure. *Mater. Today Phys.* **38**, 101233 (2023).

45 Giannozzi, P. *et al.* QUANTUM ESPRESSO: a modular and open-source software project for quantum simulations of materials. *J. Phys. Condens. Matter* **21**, 395502 (2009).

46 Zhong, X. *et al.* Prediction of Above-Room-Temperature Superconductivity in Lanthanide/Actinide Extreme Superhydrides. *J. Am. Chem. Soc.* **144**, 13394-13400 (2022).

47  Lu, S., Zhou, Q., Ouyang, Y., Guo, Y., Li, Q. & Wang, J. Accelerated discovery of stable lead-free hybrid organic-inorganic perovskites via machine learning. *Nat. Commun.* **9**, 3405 (2018).

48  Frey, N. C., Wang, J., Vega Bellido, G. I., Anasori, B., Gogotsi, Y. & Shenoy, V. B. Prediction of Synthesis of 2D Metal Carbides and Nitrides (MXenes) and Their Precursors with Positive and Unlabeled Machine Learning. *Angew. Chem. Int. Ed.* **13**, 3031-3041 (2019).

49  Heyrovska, R. Atomic, Ionic and Bohr Radii Linked via the Golden Ratio for Elements of Groups 1 - 8 including Lanthanides and Actinides. *Inter. J. Sci.* **2**, 63-68 (2013).

50  Jia, Y. Q. Crystal Radii and Effective Ionic Radii of the Rare Earth Ions. *J. Solid State Chem.* **95**, 184-187 (1991).

51  Shi, X., He, C., Pickard, C. J., Tang, C. & Zhong, J. Stochastic generation of complex crystal structures combining group and graph theory with application to carbon. *Phys. Rev. B* **97**, 014104 (2018).

52  Kresse, G. & Furthmüller, J. Efficient iterative schemes for ab initio total-energy calculations using a plane-wave basis set. *Phys. Rev. B* **54**, 11169-11186 (1996).

53  John P. Perdew, Kieron Burke & Ernzerhof, M. Generalized Gradient Approximation Made Simple. *Phys. Rev. Lett.* **77**, 3865 (1996).

54  Hamann, D. R. Optimized norm-conserving Vanderbilt pseudopotentials. *Phys. Rev. B* **88**, 085117 (2013).

55  Topsakal, M. & Wentzcovitch, R. M. Accurate projected augmented wave (PAW) datasets for rare-earth elements (RE=La–Lu). *Comput. Mater. Sci.* **95**, 263-270 (2014).

56  Dynes, R. C. McMillian's equation and the $T_c$ of superconductors. *Solid State Communications* **10**, 615-618 (1972).

57  Allen, P. B. & Dynes, R. C. Transition temperature of strong-coupled superconductors reanalyzed. *Phys. Rev. B* **12**, 905-922 (1975).

58  ELIASHBERG, G. M. Interaction between electrons and lattice vierations in a superconductor. *Sov. Phys.* **11**, 966-976 (1960).

59  Pickard, C. J. & Needs, R. J. Ab initio random structure searching. *J. Phys. Condens. Matter* **23**, 053201 (2011).

60  Clark, S. J. *et al.* First principles methods using CASTEP. *J. Phys.-Condes. Matter* **220**, 567–570 (2005).

## Acknowledgments

We thank Professor Changfeng Chen for a critical reading and extensive refinement of our paper. This work was supported by National Key R&D Program of China (Nos. 2022YFA1402304 and 2023YFA1406200), National Natural Science Foundation of China (Grants Nos. 12122405, 12274169, and 52072188), Program for Science, Technology Innovation Team in Zhejiang (Grant No. 2021R01004), the Fundamental Research Funds for the Central Universities and Graduate Innovation Fund of Jilin University (No. 2024KC047). Some of the calculations were performed at the High Performance Computing Center of Jilin University and using TianHe-1(A) at the National Supercomputer Center in Tianjin.

## Author contributions

D. D. initiated the project. T.M., D.A., S.W. and Z.Z. performed the calculations. T.M. and D.D. analyzed

data. All authors contributed to discussing the results and writing the paper.

## Competing interests

The authors declare no competing interests.

## Additional information

Supplementary Information accompanies this paper at SM.pdf.

## Data availability

The authors declare that the main data supporting the findings of this study are contained within the paper and its associated Supplementary Information. All other relevant data are available from the corresponding author upon reasonable request.